\documentclass[12pt]{article}
\usepackage[a4paper]{geometry}
\usepackage{enumerate}
\usepackage{mathtools}
\usepackage{amsmath}
\usepackage{graphicx}
\usepackage{epsfig}
\usepackage{sidecap}
\usepackage{enumerate}
\usepackage{hyperref,stackrel}
\usepackage[normalem]{ulem}
\usepackage[bottom]{footmisc}
\usepackage{color}
\usepackage{enumerate}
\usepackage{bm}
\usepackage[utf8]{inputenc}
\usepackage{amsmath}
\usepackage{amsfonts}
\usepackage{amssymb}
\usepackage{graphicx}
\usepackage{enumitem}
\usepackage{authblk}
\usepackage{hyperref}
\usepackage [english]{babel}
\usepackage{color}
\usepackage[autostyle]{csquotes}
\MakeOuterQuote{"}
\usepackage{xcolor}

\begin{document}
\title{Experimental implementation of distributed phase reference quantum key distribution protocols}

\author[1]{Satish Kumar \thanks{mr.satishseth@gmail.com}}
\author[1]{Priya Malpani \thanks{priya.ims07@gmail.com; entire work was done while she was in Jaypee Institute of Information Technology, but she has moved to C-DOT, Delhi after completion of the present work}}
\author[1]{Britant \thanks{britant808@gmail.com }}
\author[1]{Sandeep Mishra\thanks{sandeep.mtec@gmail.com}}
\author[1,@]{ Anirban Pathak\thanks{anirban.pathak@gmail.com}}
\affil[1]{Department of Physics and Materials Science \& Engineering, Jaypee Institute of Information Technology, A 10, Sector 62, Noida,
UP-201309, India}
\affil[@]{Corresponding author:anirban.pathak@gmail.com}

\maketitle
\begin{abstract}
Quantum cryptography is now considered as a promising technology due to its promise of unconditional security. In recent years, rigorous work is being done for the experimental realization of quantum key distribution (QKD) protocols to realize secure networks. Among various QKD protocols, coherent one way and differential phase shift QKD protocols have undergone rapid experimental developments due to the ease of experimental implementations with the present available technology. In this work, we have experimentally realized  optical fiber based coherent one way and differential phase shift QKD protocols at telecom wavelength. Both protocols belong to a class of protocols named as distributed phase reference protocol in which weak coherent pulses are used to encode the information. Further, we have analysed the key rates with respect to different parameters such distance, disclose rate, compression ratio and detector dead time. 
\end{abstract}


\section{Introduction}

Since the beginning of civilization, humans have been designing different
ways for sharing sensitive information. The idea is to have
information remaining private for the legitimate parties while illegitimate
parties get no information. In fact, as of now this field of cryptography
is the backbone on which secure transmissions or transactions are
performed in the current digital world. However, the currently used cryptosystems
based on mathematical complexity are only conditionally secure \cite{rothe2005complexity}. So,
the world is now looking to explore systems that promise unconditional
security \cite{mayers2001unconditional}. One of the ways forward is to exploit the features of quantum
systems to develop systems that offer unconditional security based
on the laws of physics. First, such quantum cryptography protocol was proposed in 1984
by Bennett and Brassard \cite{BB84} and since then the field has grown leaps and bounds \cite{gisin2002quantum,pirandola2020advances,shenoy2017quantum}. With the
rapidly growing technology, the quantum key distribution (QKD) has
now come out of the shadows of theoretical proofs to experimental
realization with the enhancement of key rates and distances. In fact,
the field trials of QKD systems can be traced back to the last decade
when it was tested on fiber-based networks in metropolitan areas \cite{peev2009secoqc,sasaki2011field,wang2014field,tang2016measurement} of
a few countries. Currently, some groundbreaking results have been reported
for satellite-based quantum communications, too \cite{bedington2017progress,liao2017satellite,dai2020towards}.
In fact, the field of quantum cryptography is looking very promising with many
leading corporate organizations as well as governmental organizations working
on commercial QKD systems \cite{stanley2022recent}. 

\begin{figure}[h!]
\centering{}\includegraphics[scale=0.6]{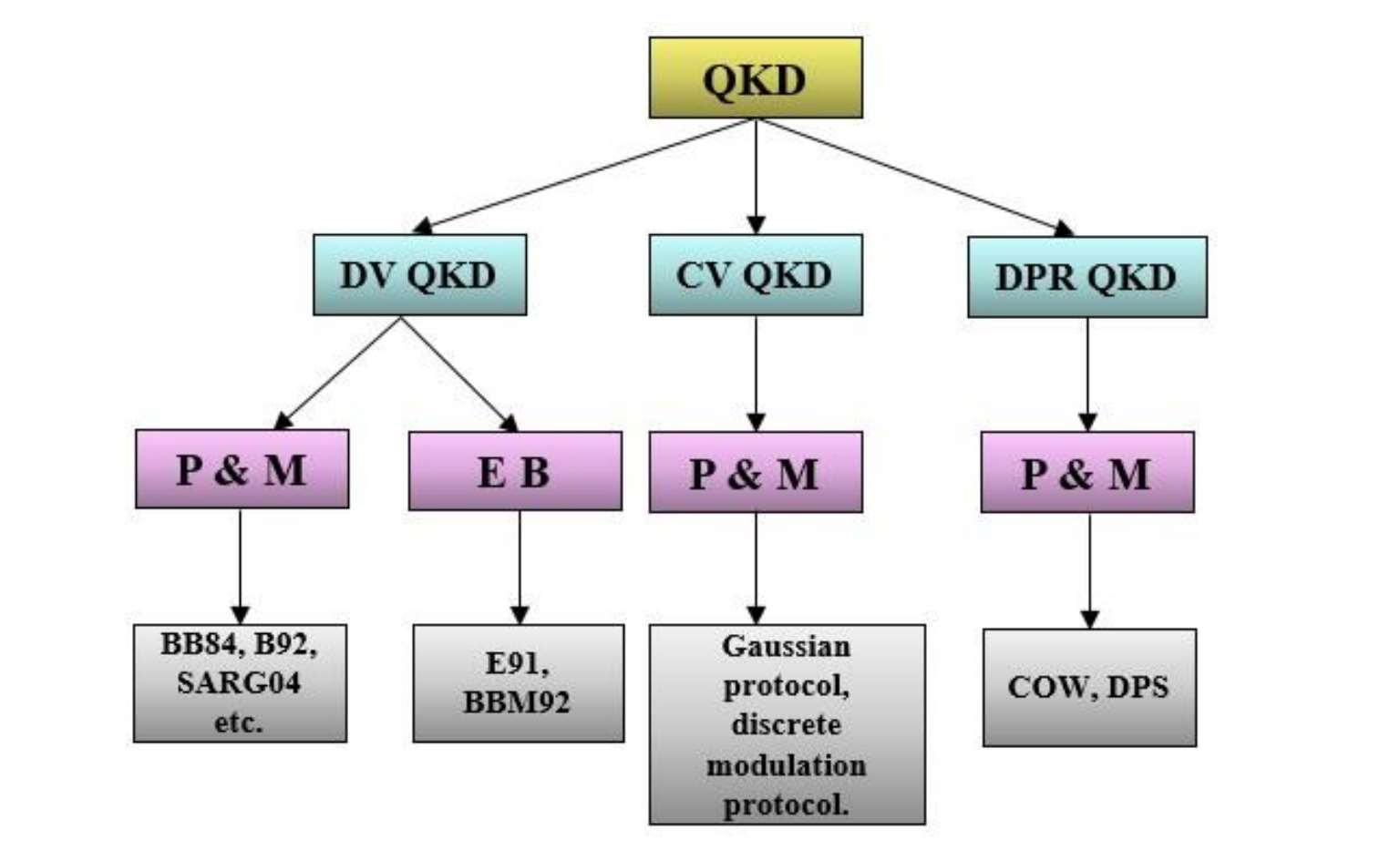}\caption{\label{fig:QKD_structure} (Color Online) Classification of QKD where
P \& M is prepare and measure based and E B is entanglement based
QKD protocol \cite{sharma2021quantum}.}
\end{figure}

If we look into the QKD schemes, then the existing protocols can be
divided into three main categories: (i) discrete variable QKD (DV QKD)
protocols \cite{BB84,bennett1992quantum,scarani2004quantum,ekert1991quantum}, 
(ii) continuous variable QKD (CV QKD) protocols \cite{ralph1999continuous},
and (iii) distributed phase reference QKD (DPR QKD) protocols \cite{inoue2002differential,stucki2005fast}
(see Fig. \ref{fig:QKD_structure}). In DV QKD protocols, the encoding
is done in discrete variables of a quantum state like the polarization
of single photons with examples being BB84 \cite{BB84}, B92\cite{bennett1992quantumm}, SARG04 protocol \cite{scarani2004quantum}. In CV QKD, the message is encoded in continuous variables like quadratures
of coherent or squeezed states with examples such as Gaussian protocol
\cite{cerf2001quantum}, Discrete modulation protocol \cite{hillery2000quantum}, CV-B92 protocol \cite{srikara2020continuous} etc. In DPR QKD, the phase difference between two successive pulses or
the arrival times of the photons are used to encode the key information.
In DPR protocols, single photon is not required for encoding, in fact,
we use weak coherent pulses (WCP). With the present available technology,
a DPR scheme is considered to be one of the most practical QKD solutions. Compared to other
QKD schemes, DPR QKD protocols have relatively easy-to-implement experimental setups and high communication efficiency. DPR QKD protocols are mainly divided
into two protocols namely: differential phase shift (DPS) QKD and
coherent one-way (COW) QKD. DPS protocol was first introduced in 2002
\cite{inoue2002differential} and COW protocol \cite{stucki2005fast} was introduced
in 2005 and since then we have seen several advances with respect to their experimental implementations
for various distances. A detailed analysis of the major advancements in the distances of DPR QKD will be discussed during the later part of this article.

Despite several successful implementations of the DPR QKD systems over various places, several issues are left unaddressed as mostly the focus revolves around the key rates and distances. During the process, several important parameters such as detector dead time (DT), disclose rate (DR), and compression ratio (CR) are often ignored. It is here to be mentioned for any practical QKD systems the key rates over a range of distances are very much dependent on these parameters. The purpose of this paper is to analyze the performance of the COW and the DPS QKD protocols for various parameters like distance, disclose rate (DR), compression ratio (CR), and detector dead time (DT). In fact, in our earlier work \cite{malpani2023}, we have reported the experimental implementation of COW QKD and made a detailed analysis of the key rates over a range of distances with respect to the parameters such as DR, CR and DT. So, in continuation with the earlier work, we will perform a similar analysis for DPS QKD. Further, it is important to mention that our implementation of COW QKD in ref \cite{malpani2023} was without the monitoring line. So in this work, we have implemented the COW QKD along with the monitoring line. With this addition of the monitoring line COW QKD,  a comparison of DPS QKD scheme with COW QKD scheme is also performed. 

The rest of the paper is structured as follows. In Section \ref{sec:COW-protocol-definition},
COW and DPS QKD protocols are discussed. In Section \ref{sec:Advances}, experimental progress regarding DPS and COW QKD schemes is discussed. The analysis of key rates with respect to different parameters is presented in Section \ref{sec:Observation}. Finally, the paper is concluded in Section \ref{sec:Conclusion}.

\section{DPS and COW QKD protocols \label{sec:COW-protocol-definition}}

DPR QKD protocols are mainly of two types namely: DPS QKD and COW QKD.
There are some similarities and some differences in both protocols
which have been highlighted in Table \ref{tab:COW-DPS}. In DPS QKD,
the encoding is done in the form of the phase difference between two consecutive
coherent pulses and the mean number of photon should be $(\mu=\left|\alpha\right|^{2}=0.2)$,
while in COW QKD, the encoding is done by combining vacuum and coherent
pulse with mean photon number $(\mu=\left|\alpha\right|^{2}=0.5)$.
Both of the protocols are robust with respect to photon number splitting
(PNS) attack and polarization sensitivity. In case of DPS QKD, the minimum
number of detectors required is $2$ whereas for COW QKD implementation,
$3$ detectors are required (one for data line and the other two are for
monitoring line).\textcolor{blue}{{} }\footnote{In our case, we have used $2$ detectors to implement COW QKD. ${\rm D}_{M2}$
clicks only when Eve is present (coherence is broken) so if Eve is
present then counts at ${\rm D}_{M1}$ will reduce. }

\begin{table}
\begin{centering}
\begin{tabular}{|p{1.5cm}|p{3cm}|p{4cm}|p{4cm}|}
\hline 
S No. & Properties & COW & DPS\tabularnewline
\hline 
1 & Encoding & Combining vacuum and coherent pulse & Phase difference between consecutive coherent pulse\tabularnewline
\hline 
2 & Source & WCP & WCP\tabularnewline
\hline 
3 & $\mu$ & $0.5$ & $0.2$\tabularnewline
\hline 
4 & PNS attack effect & No & No\tabularnewline
\hline 
5 & No of detectors & $3$ & $2$\tabularnewline
\hline 
6 & Phase & Constant & Modulated\tabularnewline
\hline 
7 & Intensity & Modulated & Constant\tabularnewline
\hline 
8 & Polarization & Insensitive & Insensitive\tabularnewline
\hline 
\end{tabular}
\par\end{centering}
\caption{\label{tab:COW-DPS} Comparison of COW and DPS QKD protocol.}
\end{table}
Now, we will briefly describe the steps involved for the implementation
of DPS QKD and COW QKD protocols.

\subsection{DPS QKD}

\begin{figure}
\begin{centering}
\begin{tabular}{c}
\includegraphics[scale=0.5]{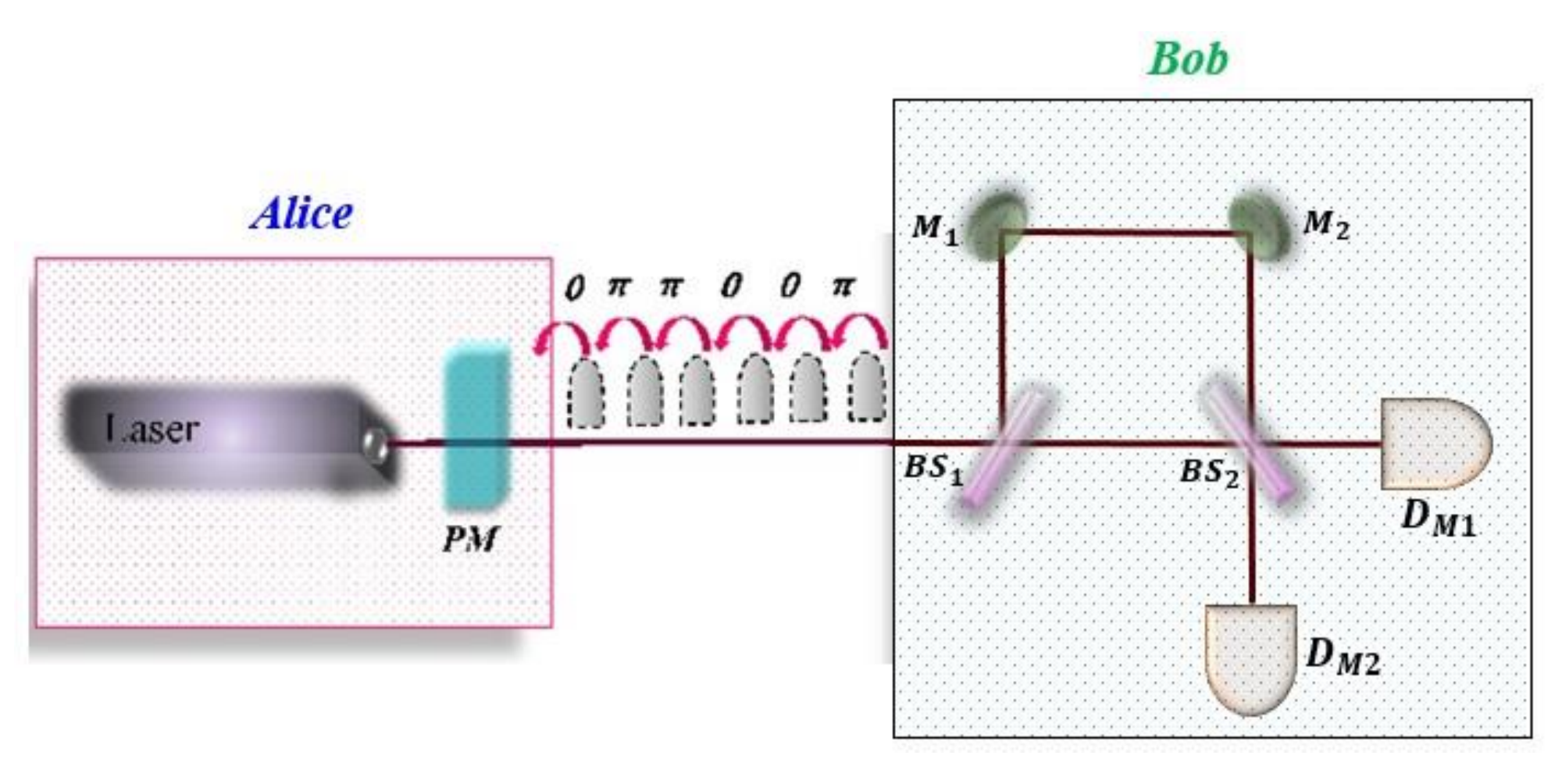}\tabularnewline
\end{tabular}
\par\end{centering}
\caption{\label{fig:DPS}(Color Online) Illustration of DPS protocol. PM: phase
modulator, ${\rm BS_{1}}$, ${\rm BS_{2}}$ are beamsplitters and
${\rm M_{1}}$, ${\rm M_{2}}$ are mirrors.}
\end{figure}

The DPS QKD protocol was proposed in 2002 by Inoue et al. \cite{inoue2002differential}.
In this protocol, Alice creates a weak coherent pulse (WCP) whose average
photon number is $0.2$ and randomly modulates the phase either by
$0$ or $\pi$ using a phase modulator. She sends these modulated
pulses to Bob using an optical fiber. At the receiver's side, Bob passes
the received pulses to a $1$ bit delay Mach Zehnder interferometer.
The outputs of the Mach Zehnder interferometer are measured either
by single photon detectors (SPDs) or superconducting nanowire single
photon detectors (SNSPDs). The two consecutive pulses interfere with
each other in the Mach Zehnder interferometer with the detection from
either of the two detectors having information of the phase difference
between between the two consecutive pulses. In Fig. \ref{fig:DPS},
click on $D_{M1}\,\left(D_{M2}\right)$ means phase difference $0\,(\pi)$
between the two successive pulses. The steps involved in the DPS QKD protocol
are as follows:
\begin{enumerate}
\item Alice sends a random sequence of phase-modulated $(0,\pi)$ attenuated
pulses to Bob, with each pulse having an average photon number less than
one.
\item Bob uses photon detectors and a $1$ bit delay Mach Zehnder interferometer
to measure the phase difference between two consecutive pulses and
records the information of detector clicks as well as time of arrival
of photons. 
\item Bob publicly informs Alice about the times when either of the detectors
clicked.
\item With the timing information about Bob's detectors clicks, Alice obtains
`which detector clicks' information at Bob's side.
\item Key bits $0$ and $1$ are respectively assigned if the phase difference
between two successive pulses are $0$ or $\pi$. 
\item Alice and Bob perform error correction and privacy amplification on
the sifted key to obtain the secret key.
\end{enumerate}

\subsection{COW QKD}

\begin{figure}
\begin{centering}
\begin{tabular}{c}
\includegraphics[scale=0.5]{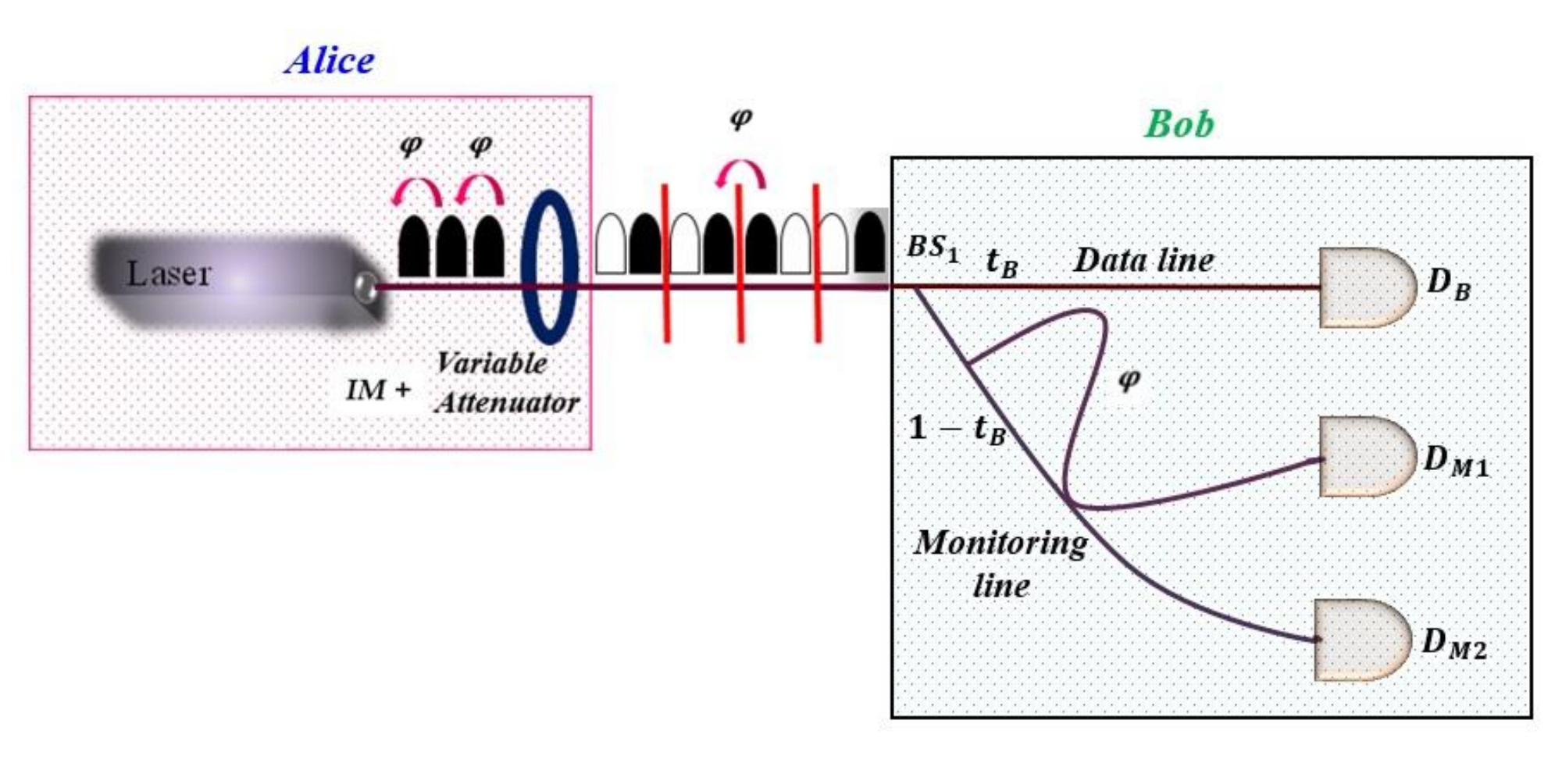}\tabularnewline
\end{tabular}
\par\end{centering}
\caption{\label{fig:COW}(Color Online) Illustration of COW protocol. IM: intensity modulator, BS1: beamsplitter }
\end{figure}

In COW QKD protocol, the encoding of a bit is done using a pair of
empty and non-empty pulse. It was first introduced by Stuki et al.
\cite{stucki2005fast,stucki2009continuous} and the steps involved in the protocol
are as follows: 
\begin{enumerate}
\item Alice prepares a sequence of pulses $\left|0\right\rangle \left|\alpha\right\rangle $
(empty, non-empty), $\left|\alpha\right\rangle \left|0\right\rangle $
(non-empty, empty) and $\left|\alpha\right\rangle \left|\alpha\right\rangle $
(non-empty, non-empty) $\left(\left|\alpha\right|^{2}<1\right)$ corresponding
to logical bit $1$, $0$ and decoy respectively using attenuated
light source and intensity modulator with each logical bit having
probability $\frac{\left(1-f\right)}{2}$ with the decoy occurring
with probability $f$ . Alice sends these pulse sequence to Bob via
high quality optical fiber.
\item Bob receives the pulse sequence and measures the time of arrival of
$90\%$ photons on his detector $D_{{\rm B}}$ for the generation
of sifted key while the rest $10\%$ of photons are measured on the
monitoring line for security purpose (refer to Fig. \ref{fig:COW}).\footnote{$90:10$ distribution of photons among data line and monitor line
mentioned here and implemented by us is standard. However, some groups
have implemented with $95:5$ distribution \cite{shaw2022optimal}.
Such an approach would lead to higher KR at the cost of security as
the number of pulses in monitor line will considerably reduces leading
to a possibility that an eavesdropping effect remains unnoticed.} 
\item Bob randomly checks the coherence between the non-empty pulses using the 
detector $D_{{\rm M1}}$ and $D_{{\rm M2}}$ in the monitoring line.
Basically, the monitoring line is a Mach Zehnder interferometer arranged
in such a manner that detector $D_{{\rm M1}}$ will click if there
is no disturbance by Eve (refer Fig. \ref{fig:COW}). Alice and Bob
abort the protocol if the number of detection at $D_{{\rm M2}}$ is
more than the threshold level.
\item Alice and Bob generate a sifted key from the pulses received from
the data line. The sifted key undergoes through error correction and
privacy amplification to get the private key which can be used for
encryption and decryption.
\end{enumerate}

\subsection{Postprocessing}

The method described above for COW QKD and DPS QKD is used by the sender and the receiver to respectively encode and decode the raw key. This raw key cannot be used for practical applications. In fact, the key sent by the sender and receiver may not be exactly the same as some errors may have been introduced due to the imperfectness of the devices and the losses in transmission channel. Moreover, one also needs to check for the presence of Eve during the transmission of the keys. So, postprocessing of the raw key is required in order to check for the presence of Eve as well as to correct the errors introduced during the transmission. The key generated after following postprocessing schemes is then used for practical applications. The major steps involved in post processing as as follows:

\begin{enumerate}
\item {\bf Parameter estimation:} The idea of the parameter estimation is to look for quantum bit error rate (QBER). Since any device and the transmission channel will never be perfect, so some errors will always be present in the raw key. Many a time Eve would also be listening through the channel in order to get the information about the key. So, it is always important to distinguish between the errors introduced due to interception by Eve and the naturally occurring errors due to device and channel imperfections. So, the sender and receiver will publicly announce some subset of their respective raw keys and will estimate the QBER. The fraction of raw key used for the estimation of QBER is known as disclose rate (DR). A higher value of QBER is not desired as Eve can hide behind the errors to get the information about the key. If the QBER is above a particular threshold, then the communication will be aborted. Since the estimation of QBER is based on the DR, so a smaller value of DR can lead to wrong conclusions for QBER. Ideally, the DR should be 50 percent, but such a high value of DR will lead to a decrease in  key rates. So, for the cases of large raw key and randomly selected sample bits used for estimation of QBER, DR in the range of 3-10 percent \cite{israel1992determining} is statistically sufficient to detect the presence of Eve.  

\item {\bf Error correction:} In this step, the sender and receiver will identify the locations in a raw key where errors are introduced and then follow some protocols to correct those errors. Usually, the process of error correction is based on the estimation of QBER. There are many error correction schemes, but low-density parity-check (LDPC) scheme \cite{mackay1999good} in mostly used in currently available commercial QKD systems. The error correction is a classical process that employs some exchange of information over the classical channel. After the completion of the process, both the sender and receiver will have identical keys.

\item {\bf Privacy amplification:} Since, the error correction step involves the exchange of information over a classical channel, so Eve may listen to the conversation and can get some partial information about the key. So, the process of privacy amplification is to shorten the key and thus reduce Eve's information of the key to a very negligible level. The factor with which the key is shortened is known as  compression ratio (CR). There are various privacy amplification schemes available  which can be used. 

\end{enumerate}
So, the final key rate (KR) achieved after the post processing process can be expressed as:
\begin{eqnarray*}
\text{KR} & = & \text{Raw Key Rate \ensuremath{\times\text{(1-DR)\ensuremath{\times\text{(1-CR)}}}}}
\end{eqnarray*}

\section{Experimental implementation of DPR QKD protocols and comparison of key rates with other contemporary implementations\label{sec:Advances}}

The DPS QKD protocol was proposed in 2002 by Inoue et al. \cite{inoue2002differential} while the COW QKD was introduced by Stuki et al \cite{stucki2009continuous}. During the last 20 years, there has been a tremendous amount of advancements with ever increasing key rates and distances. In this section, along with our experimental realization of the DPS QKD  scheme, we would also like to mention some of the other important experimental realizations of DPS QKD and COW QKD  with Table \ref{tab:Literature survey}
(\ref{tab:Literature survey-1}) summarizing the progress made for DPS (COW) QKD. 
\begin{table}
\begin{centering}
\begin{tabular}{|p{1.5cm}|p{1.5cm}|p{2.5cm}|p{1.5cm}|p{1.5cm}|p{2cm}|}
\hline 
QKD & Mode & Detector Type & Distance (km) & KR (bps) & References\tabularnewline
\hline 
DPS & Fiber &   APD & $80$ & $21352$ & our work\tabularnewline \hline
DPS & Fiber{*} &  up-conversion-assisted hybrid photon detector (HPD) & $10$ & $1.3$Mbps & \cite{zhang2009megabits}\tabularnewline
\hline 
DPS & Fiber &  SNSPD & $90$ & $2100$bps & \cite{sasaki2011field}\tabularnewline
\hline
DPS & Fiber &  APD & $20$ & $3076$ & \cite{honjo2004differential}\tabularnewline
\hline 
DPS & Fiber & InGaAs APD & $25$ & $9000$ & \cite{honjo2009differential}\tabularnewline
\hline 
DPS & Fiber &  Low jitter up-conversion detector & $100$ & $106$ & \cite{diamanti2006100}\tabularnewline
\hline 
DPS & Fiber &  low-jitter up-conversion detector
$10$ GHz clock & $105$ & $3700$ & \cite{takesue200610}\tabularnewline
\hline 
DPS & Fiber &  InGaAs/InP APD ($6\%$ eff) & $100-160$ & $24000-490$ & \cite{namekata2011high}\tabularnewline
\hline 
DPS & Fiber & SNSPD & $200$ & $12.1$ & \cite{takesue2007quantum}\tabularnewline
\hline 
DPS & Fiber &  Ultra low noise SNSPD & $250$ & \# & \cite{wang20122}\tabularnewline
\hline 
DPS & Fiber{*} & SNSPD with ultra low DCR $0.01$counts per second & $336$  & \# & \cite{shibata2014quantum}\tabularnewline
\hline 
DPS & Fiber & SNSPD  with $0.01$counts per second & $265$  & $192$ & \cite{pathak2023phase}\tabularnewline
\hline
\end{tabular}
\par\end{centering}
\caption{\label{tab:Literature survey}Table represents the
developments in DPS QKD protocols. \# means that the information
is not given in the mentioned paper. {*} is for dispersion shifted.
fiber}
\end{table}

\begin{table}
\begin{centering}
\begin{tabular}{|p{1.5cm}|p{1.5cm}|p{2.5cm}|p{1.5cm}|p{1.5cm}|p{2cm}|}
\hline 
QKD & Mode  & Detector Type & Distance (km) & KR (bps) & References\tabularnewline
\hline 
COW & Fiber & InGaAs SPD (free running mode) & (a)$120$

(b) $145$ & (a) $241-2410$

(b)$154-1184$ & \cite{malpani2023}\tabularnewline
\hline 
COW & Fiber & InGaAs SPD (Lab) & $150$ & $>50$ & \cite{stucki2009continuous}\tabularnewline
\hline 
COW & Fiber & SNSPD (Field) & $150$ & $2.5$ & \cite{stucki2009continuous}\tabularnewline
\hline 
COW & Fiber & SNSPD & $100-250$ & $6000-15$ & \cite{Gisin2009COW}\tabularnewline
\hline 
COW & Fiber & InGaAs SPD & (a) $14.2$

(b) $45.6$

(c) $24.2$

(d) $36.6$

(total 121 km with three trusted nodes) & (a) $1956$

(b) $1314$

(c) $763$

(d) $1906$ & \cite{wonfor2019field}\tabularnewline
\hline 
COW & Fiber & InGaAs/InP negative feedback avalanche diodes (NFADs) & $307$ & $3.18$ & \cite{korzh2015COW}\tabularnewline
\hline 
COW & Fiber & InGaAs SPD & $25$ & $22500$ & \cite{walenta2014COW}\tabularnewline
\hline 
COW & Fiber & InGaAs SPD (gated mode) & $150$ & $<1000$ & \cite{shaw2022optimal}\tabularnewline
\hline 
\end{tabular}
\par\end{centering}
\caption{\label{tab:Literature survey-1} Experimental developments for the
fiber based COW-QKD protocol. \# means that the information is not
provided clearly in the mentioned paper.}
\end{table}
We have realized both the COW QKD and DPS QKD protocol in cooperation with Centre for Development of Telematics (C-DOT) \cite{Cdot} in a rack mountable in four $19$ inches boxes. In this rack, both Alice's and Bob's station are there but are separated by spools of optical fibre. The distance between the Alice and Bob can be varied by altering the length of the optical fibre spools. Further, the parameters such as DR, CR and DT are varied to look for the corresponding changes in the key rates. It is worth mentioning that we have already reported the experimental realization of COW QKD in our earlier work and analysed the variation of key rates with different parameters such as DR, DT, CR over a range of distances \cite{malpani2023}. But, the experimental set-up didn't have the monitoring line as it was demonstrated in lab alone where we were sure about the absence of an Eve. But in this article, we have analysed the COW QKD set-up along with the monitoring line also. The specifications of the components used in our experiment has been tabulated in Table \ref{tab:components used}. 
\begin{table}
\begin{centering}
\begin{tabular}{|c|c|}
\hline 
\textbf{Component/ Technique} & \textbf{Property/ Type}\tabularnewline
\hline 
Laser & PS-NLL laser from Teraxion\tabularnewline
\hline 
Fiber & SMF-28 (ITU-TG652D) (loss-0.2 dB/km)\tabularnewline
\hline 
Intensity modulator (COW) & Lithium Niobate based\tabularnewline
\hline 
Phase modulator (DPS) & MPZ-LN series\tabularnewline
\hline 
Random number generator & TRNG\tabularnewline
\hline 
Operating Temperature & $ 10^{o}{\rm C}-2\ensuremath{8^{o}}{\rm C}$\tabularnewline
\hline 
Detector & SPD\_OEM\_NIR from Aurea Technology\tabularnewline
\hline 
Error correction technique & LDPC\tabularnewline
\hline 
Privacy amplification & Toeplitz based hashing\tabularnewline
\hline 
\end{tabular}
\par\end{centering}
\caption{\label{tab:components used} Specifications of the
components used in our experiment.}
\end{table}

In the experiment that we have performed with DPS QKD set-up, we have altered the fibre spool to generate keys over distances of $80$ km, $100$ km and $120$ km over a range of different values for the parameters DR, CR and DT. The highest key rate that we have been able to generate is $21352$ bps for $80$
km with  DR $=3.125\%$, CR $=50\%$, DT $=20\,\mu$s  and dark count rate of approx  800 counts per second  \cite{Aurea}). In fact, the key rates generated are in sync with the comparative experiments performed by other groups. e.g. Namekata et al. in 2011 reported KR of $24$ kbps for $100$ km. The reason for them getting a slightly better  KR is the use of APD in gated mode and the use of dispersion-shifted fiber (DSF). They have used $2$-GHz sinusoidally gated APDs which has a dark count probability of $2.8\times10^{-8}$ ($55$ counts per second) and has shown that sinusoidally gated APDs have exceeded the performance of the SNSPD. During the field trials at Tokyo in 2011, the group led by Sasaki et al. achieved an average key rate of 2.1 Kbps using a superconducting nanowire single-photon detector (SNSPD) over standard telecom fiber \cite{sasaki2011field}.  Further, it has been shown that via the use of SNSPD and the use of ultra-low loss fibre the key rates can be enhanced. e.g Wang et al. \cite{wang20122} have experimentally demonstrated the QKD over $260$ km standard telecom fibre via the use of ultra-low loss fiber with a loss coefficient $0.164$ dB/km. Similarly, in our case also the KR can be improved with the use of ultra-low loss fibre or we can increase the distance over which the QKD can be implemented. In another important experiment, Shibata et al. \cite{shibata2014quantum} via  the use of SNSPD and ultra low loss optical fiber achieved QKD over a distance of $336$ km  with key rate of 0.03 bps. Recently, Pathak et al. implemented the DPS QKD scheme over a distance of $265$ km with a secure key rate of $193$ bps and QBER of less than one percent \cite{pathak2023phase}. The experiment has been performed over standard telecom-grade optical fiber and use of SNSPD. Further, they have extrapolated the conditions to show that via the use of ultra-low loss optical fiber, the distance can be increased to $380$ km with secure key rates of $0.11$ bps. We wish to emphasize once again that via use of standard telecom grade optical fibre and use of APD we have achieved the key rates in the range of $10-25$ kbps over a distance range of $80-120$ km. The key rates in our experiment are in sync with those achieved by other experimental groups working in the field. Further, we wish to mention that the key rates are dependent upon different parameters such as DR, CR and DT. Moreover, the parameters such as DR, CR and DT are rarely discussed in most of the previously reported works. So, in the next section, we will analyse the effect of such parameters on the key rates.

\section{Analysis of key rates with different parameters\label{sec:Observation}}

In this section, we have performed an analysis of key rates with respect to different parameters over a range of distances for the case of DPS QKD. Further, we have performed a comparative analysis of the key rates for DPS QKD and COW QKD. 

\subsection{Key rate variation of DPS with disclose rate and compression ratio}

\begin{figure}
\begin{centering}
\begin{tabular}{ccc}
\includegraphics[scale=0.5]{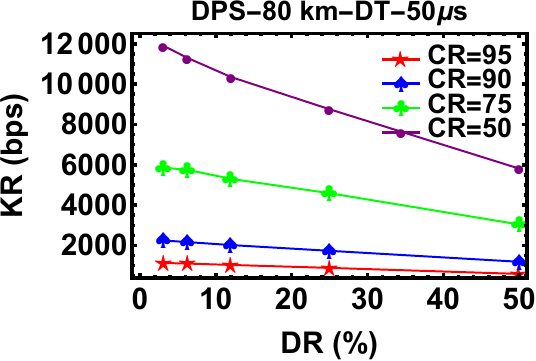} & \includegraphics[scale=0.5]{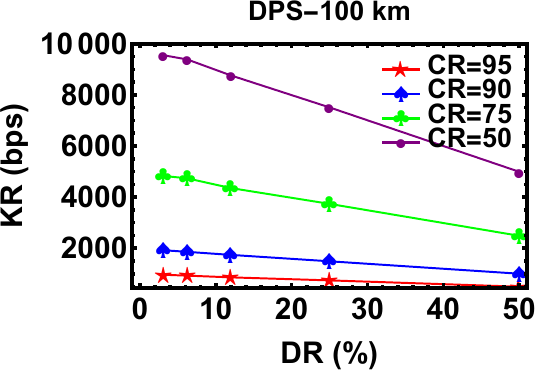} & \includegraphics[scale=0.5]{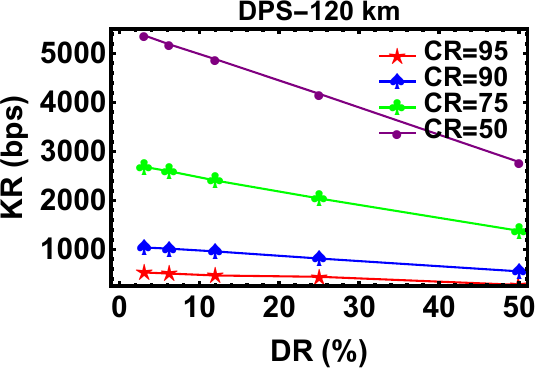}\tabularnewline
(a) & (b) & (c)\tabularnewline
\includegraphics[scale=0.5]{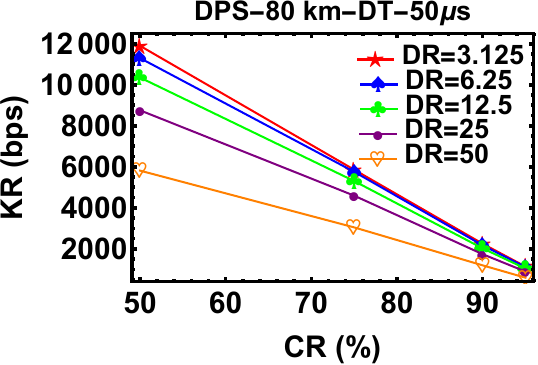} & \includegraphics[scale=0.5]{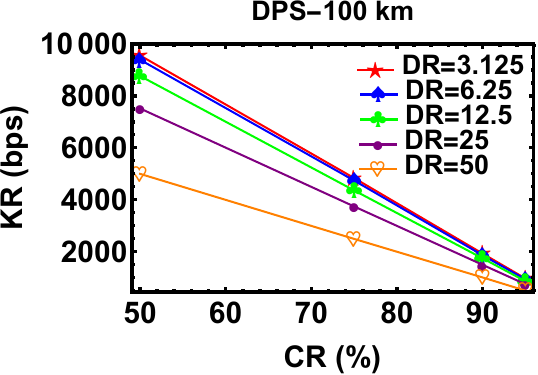} & \includegraphics[scale=0.5]{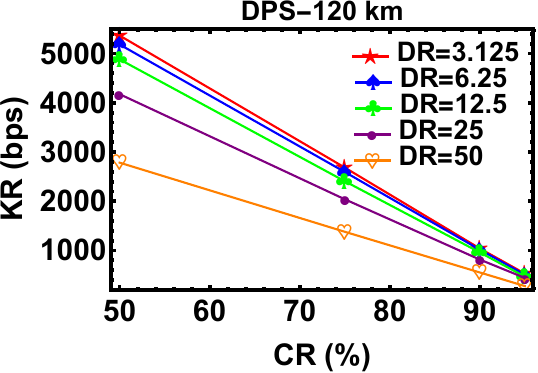}\tabularnewline
(d) & (e) & (f)\tabularnewline
\end{tabular}
\par\end{centering}
\caption{\label{fig:DPS-KR-CR} (Color Online) The KR for DPS protocol with
respect to CR for different DR (a) for $80$ km (b) for $100$ km,
(c) for $120$ keeping DT $=50\,\mu$s. KR is decreasing with increase
in the distance and increasing with DR. }
\end{figure}

We varied the distance between Alice and Bob by attaching fibre spools of length $80$ km, $100$ km and $120$ km. Then for a particular value of distance we varied the parameters of DR and CR after keeping the value of DT fixed at $50\,\mu$s. The plots shown in Fig. \ref{fig:DPS-KR-CR} (a), (b) and (c) are
for KR as a function of DR at different CR for distances $80$, $100$
and $120$ km respectively keeping DT $50\,\mu$s. The plots shown
in Fig. \ref{fig:DPS-KR-CR} (d), (e), and (f) are also for KR as
a function of CR at different DR for various distances $80$, $100$
and $120$ km respectively for DT $=50\,\mu$s. As expected, we can see  KR increases with decrease in DR and CR for various distances.

\subsection{Key rate variation of DPS with detector dead time}

\begin{figure}
\begin{centering}
\begin{tabular}{ccc}
\includegraphics[scale=0.5]{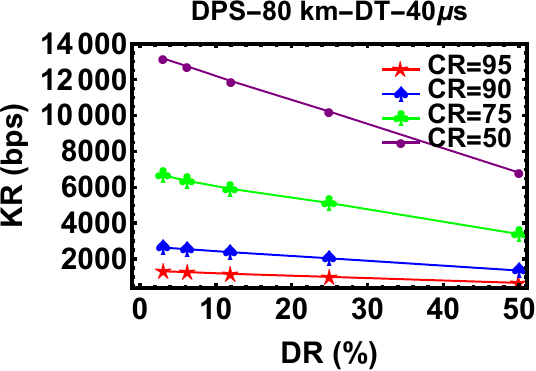} & \includegraphics[scale=0.5]{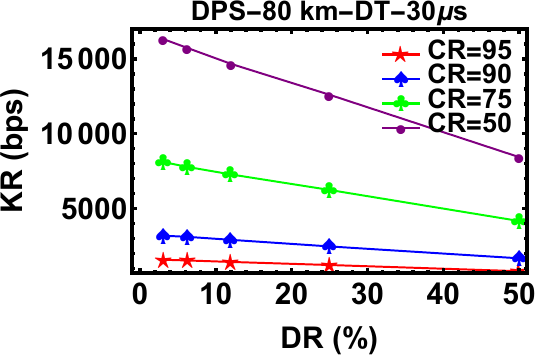} & \includegraphics[scale=0.5]{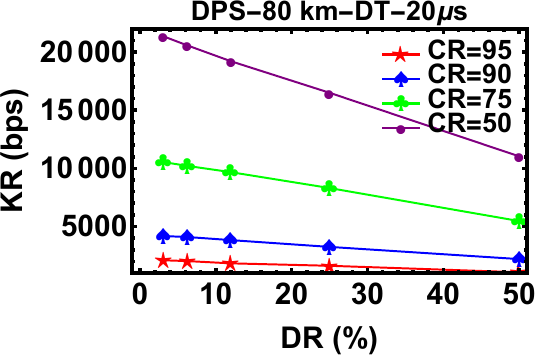}\tabularnewline
(a) & (b) & (c)\tabularnewline
\includegraphics[scale=0.5]{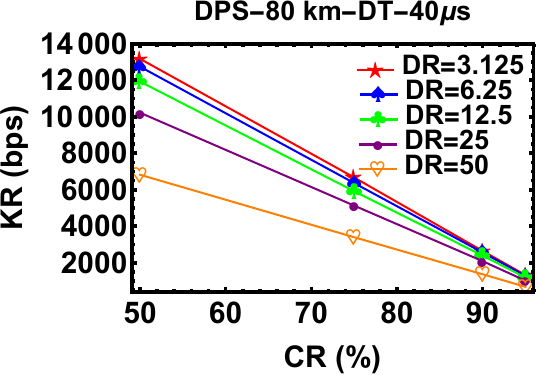} & \includegraphics[scale=0.5]{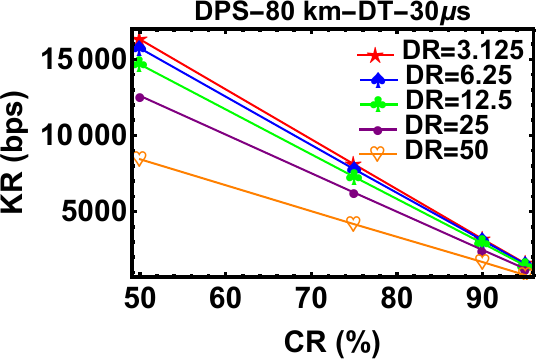} & \includegraphics[scale=0.5]{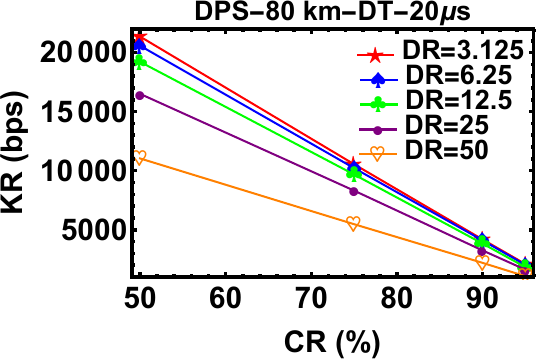}\tabularnewline
(d) & (e) & (f)\tabularnewline
\end{tabular}
\par\end{centering}
\caption{\label{fig:DPS-KR-CR-PA} (Color Online) Figure \ref{fig:DPS-KR-CR-PA}
(a)-(c) ((d)-(f)) shows the KR for DPS protocol for various DT with
respect to DR (CR) for different CR (DR) keeping quantum channel distance
$80$ km. KR increases with decrease in DT. }

\end{figure}

DT plays a vital role in all QKD related experiments. DT of a detector is defined as the time after the detection of each event during which the detector becomes inactive which means that it will not be able to record any event during that time interval. If the detector has longer DT, this means the detector won't be able to detect any photons for a longer time. So, ideally the detector dead time should be as small a possible. So, next experiment was to look for variations in key rates with respect to the decrease in DT while keeping the CR and DR constant. As expected the Figures \ref{fig:DPS-KR-CR-PA} (a)-(f), show an increase of KR with decrease in DT.

\subsection{Key rate variation of COW with disclose rate and compression ratio}

\begin{figure}
\begin{centering}
\begin{tabular}{cc}
\includegraphics[scale=0.7]{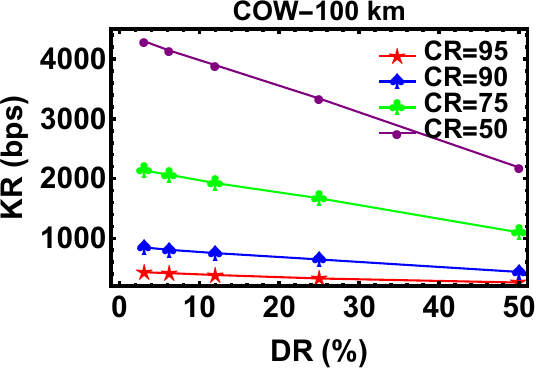} & \includegraphics[scale=0.7]{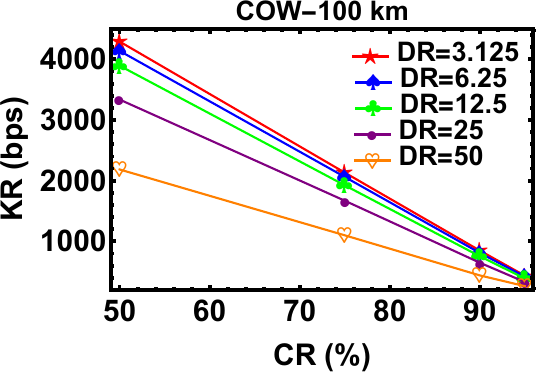}\tabularnewline
(a) & (b)\tabularnewline
\end{tabular}
\par\end{centering}
\caption{\label{fig:COW-KR-CR-PA-1} (Color Online) Figure shows the KR for
COW QKD protocol keeping $100$ km distance, DT $=50\,\mu$s, (a)
DR for various CR (b) CR for various DR.}
\end{figure}

We have mentioned before that in our earlier work \cite{malpani2023}, we have implemented the COW QKD setup without the monitoring line. So, in this work we performed the experiment again with the addition of monitoring line too. We can see in Figure \ref{fig:COW-KR-CR-PA-1} (a) and (b) the variation of  KR of COW QKD as a function of DR (CR) at different CR (DR) for fixed quantum communication distance $100$ km and fixed detector
DT which is equal to $50\,\mu$s. As observed before, the KR increases with the decrease in CR (cf. \ref{fig:COW-KR-CR-PA-1} (a)) whereas, KR decreases with the increase in DR (cf. \ref{fig:COW-KR-CR-PA-1} (b)). The similar nature of the results we have observed in our previous work as well
\cite{malpani2023}. Next, we will perform an analysis of key rates for DPS QKD with COW QKD.

\subsection{Analysis of DPS and COW key rate}

\begin{figure}
\begin{centering}
\begin{tabular}{c}
\includegraphics[scale=0.8]{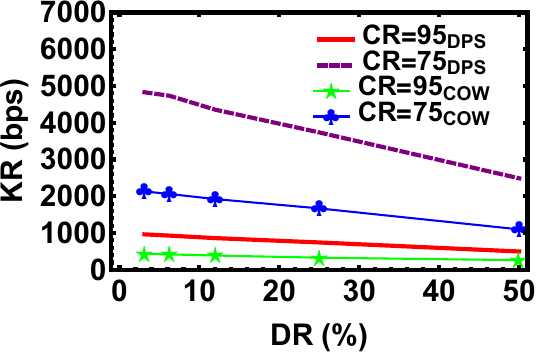}\tabularnewline
\end{tabular}
\par\end{centering}
\caption{\label{fig:COW-DPS} (Color Online) Figure shows the KR with respect
to DR for various CR for DPS (Red solid and Purple dashed line) as
well as COW (Green and Blue line with plot markers) QKD protocol.
From the figure, it is clear that keeping the same parameters except for repetition
rate ($1$ GHz in case of DPS and $500$ MHz for COW ), KR for DPS
is more than COW.}
\end{figure}

Since DPS QKD and COW QKD belong to the same class of QKD schemes so it would be apt to undertake the performance analysis of DPS QKD with COW QKD while keeping the same parameters except the repetition rate which is $1$ GHz in case of DPS QKD and $500$ MHz in case of COW QKD. As the nature of encoding in both the protocols is completely different (in case of DPS protocol all the pulses are non-empty means they do not have a vacuum part whereas in the case of COW protocol empty (vacuum) and non-empty both pulsed are used) so the repetition rate for the same is chosen accordingly in order to do a performance analysis.  We performed the experiment for a distance of  $100$ km while varying the parameters of DR and CR and keeping DT $=50\,\mu$s. Figure \ref{fig:COW-DPS} shows that KR for DPS is significantly higher as compared to COW QKD.

\section{Conclusion\label{sec:Conclusion}}

Quantum technology is a very rapidly growing technology and plays a very significant role in secure communication. QKD is now a rapidly developing field with many field trials being conducted across the world and newer corporations and governmental agencies investing heavily  in the area. QKD systems have obtained a significant level of maturity but due to some technical issues, commercial QKD system in the metropolitan area is still a challenging task. For this purpose, the requirement of low-loss fibers, cost-effectiveness and many more parameters are necessary. So many efforts are being made in this direction, one can see that DPR protocols are relatively easier to implement using the existing infrastructure. DPR QKD protocols include COW and DPS QKD protocols. Unlike, most of the quantum cryptographic tasks which are based on single photons (unfortunately, single photons are difficult to realize experimentally
on demand) DPR QKD protocols are implemented using attenuated laser pulses in which the photons obey Poissonian statistics.  In this work, we have performed a DPR QKD-based experiment, in which we implemented the DPS as well as COW QKD experiment for various distances and analysed the key rates with respect to parameters such as DR, CR and DT. The key rates obtained are in sync with the key rates obtained by other groups working in the field. Our work will help in providing some insights for the optimization of the parameters for enhancement of key rates over large distances. Further, it is one more step towards the eventual realization of secure networks.

\section*{Acknowledgements}

Authors acknowledge the support from the QUEST scheme of Interdisciplinary Cyber Physical Systems (ICPS) program of the Department of Science and Technology (DST), India (Grant No.: DST/ICPS/QuST/Theme-1/2019/14
(Q80)) and the C-DOT team working on quantum technologies. They also thank Kishore Thapliyal and Abhishek Shukla for their interest in the work.

\section*{Availability of data and materials}

Experimental data will be made available on request.

\section*{Competing interests}

The authors declare that they have no competing interests

\bibliographystyle{unsrt}
\bibliography{COW_DPS}

\end{document}